\documentclass[twocolumn,aps,floatfix,showpacs,superscriptaddress]{revtex4-1}
\usepackage{amsmath,amssymb,eucal,graphicx,float,epstopdf}

\usepackage{ulem}
\usepackage{color}

\begin{document}
\title{Reply to ``Comment on Generalized Exclusion Processes: Transport Coefficients''}

\author{Chikashi Arita}
\affiliation{Theoretische Physik, Universit\"at
 des Saarlandes, 66041 Saarbr\"ucken, Germany}
\author{P. L. Krapivsky}
\affiliation{Department of Physics, Boston University, Boston, MA 02215, USA}
\author{Kirone Mallick}
\affiliation{Institut de Physique Th\'eorique, IPhT, CEA Saclay
and URA 2306, CNRS, 91191 Gif-sur-Yvette cedex, France}

\begin{abstract}
We reply to the comment of Becker, Nelissen, Cleuren, Partoens, and Van den Broeck \cite{Com} on our article \cite{we_14} about transport properties of a class of generalized exclusion processes.
\end{abstract}


\maketitle

Stochastic lattice gases with symmetric hopping are described, on a coarse-grained level, by diffusion equation with density-dependent diffusion coefficient. Density fluctuations additionally depend on the local conductivity (which also describes the response to an infinitesimal applied field). A hydrodynamic description therefore requires the determination of these two transport coefficients. Generally for lattice gases even with rather simple hopping rules, analytic results are unattainable; however, when an additional feature, known as the {\it gradient condition}, is satisfied, the Green-Kubo formula takes a simple form \cite{Spohn} and computations of the transport coefficients become feasible. For a number of lattice gases of gradient type, e.g., for the Katz-Lebowitz-Spohn model with symmetric hopping \cite{KLS}, for repulsion processes \cite{Krapivsky}, for a lattice gas of leap-frogging particles \cite{CCGS,GK}, the diffusion coefficient has been rigorously computed. The gradient property is also true for the misanthrope process, a class of generalized exclusion processes \cite{C-T,AM}.

For gradient type lattice gases, an exact expression for the
diffusion coefficient can also be obtained by a perturbation approach: one writes
the formula for the current at the discrete lattice level and then performs
a continuous limit assuming that the density field is slowly varying.

Generalized exclusion processes with multiple occupancies \cite{KLO94,KLO95,Timo,BNCPV13}, in general, do not obey the gradient condition. However, we argued in \cite{we_14} that the perturbation approach should, nevertheless, lead to an exact prediction for the diffusion coefficient. For the class of generalized exclusion processes which we studied \cite{we_14} simulation results were indeed very close to the predictions by perturbative calculation. The comment \cite{Com} by Becker {\it et al.} 
prompted us to perform more simulations and to analyze our results more carefully. 

 \begin{figure}[b]
 \centerline{ \includegraphics[width=85mm]{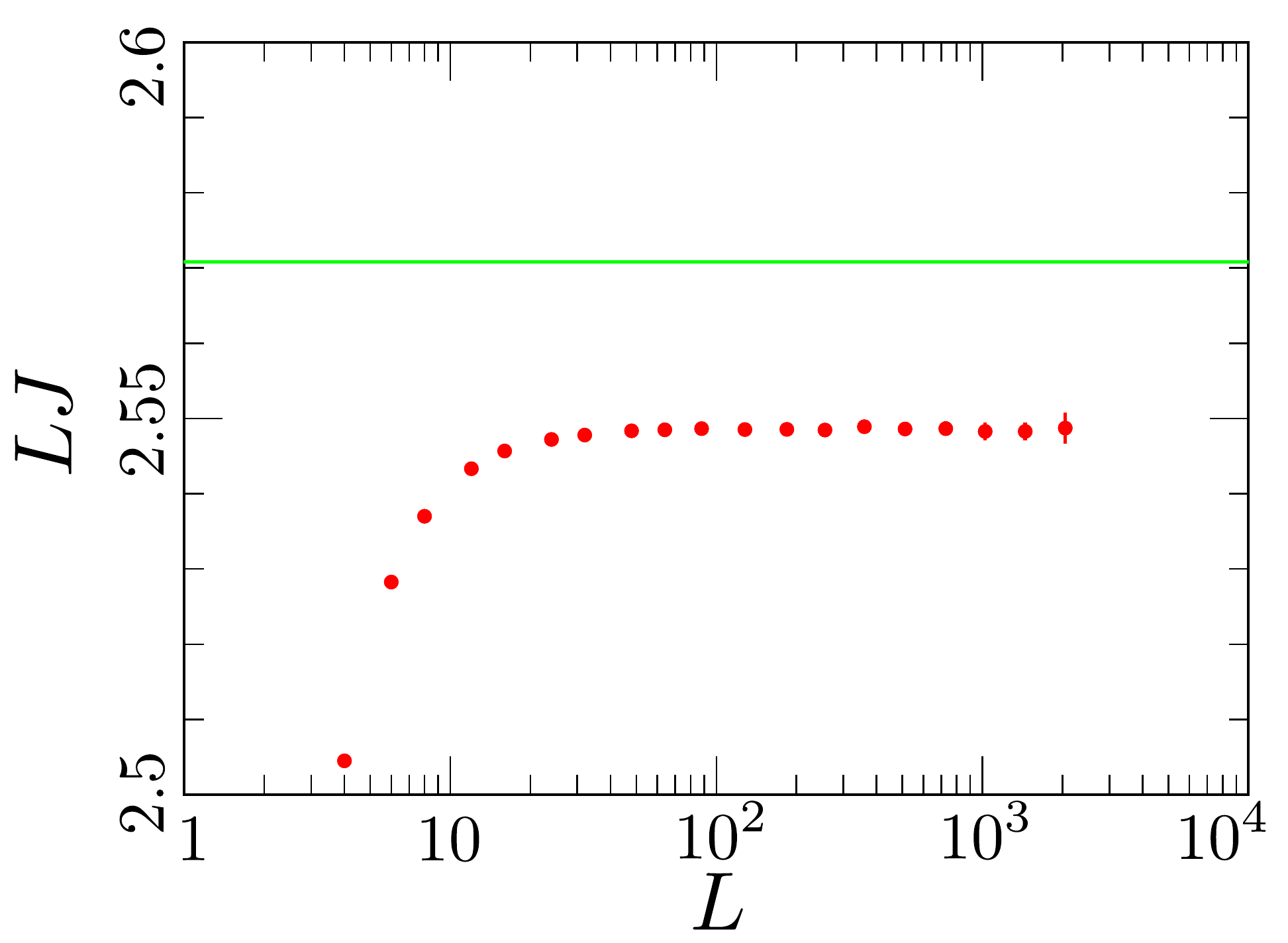} }
 \caption{Stationary current multiplied by the system size: simulation results (dots) and the prediction from our previous approach. The latter holds for $L=\infty$, but is shown as a line. 
}
 \label{fig:RP}
 \end{figure}

Becker {\it et al.} computed numerically the diffusion coefficient $D(\rho)$. They performed simulations for various system sizes $L$ and various density differences $\delta \rho$ between the boundary reservoirs. In order to extract $D(\rho)$ from simulations they needed to take \cite{Com} two limits: $L\to\infty$ and $\delta \rho\to 0$. We considered a system with a large density difference and measured the stationary current through the system: the advantage is that we have to take only one limit, $L\to\infty$. We analyzed the generalized exclusion process GEP(2) with maximal occupancy $k=2$ particles per site and extreme densities at the boundaries: $\rho(0)=2$ and $\rho(L)=0$. According to our expectations \cite{we_14}, the average current should vanish as
$(1+\frac{\pi}{2})/L$ when $L\gg 1$. Simulation results (Fig.~\ref{fig:RP}) demonstrate that the error is smaller than
$0.9\%$, but this discrepancy does not seem to disappear in the $L\to\infty$ limit.

The numerical results of Ref.~\cite{Com} and our simulations (Fig.~\ref{fig:RP}) show that the perturbation approach does not lead to the correct analytical results for the GEP(2). We emphasize that the perturbation approach is {\it not} a naive mean-field theory where correlations are obviously neglected as argued by Becker {\it et al.} In dense lattice gases, the equilibrium state itself is usually highly correlated; e.g., in the repulsion process $\langle \tau_i \tau_{i+1}\rangle =0 \ne \rho^2$ for $0\leq \rho\leq \frac{1}{2}$, where $ \tau_i \in \{1,0\} $ denotes the occupation number of site $i$: the mean-field assumption is completely wrong. Yet, a careful use of the perturbation approach leads to the correct result \cite{Krapivsky}. 

The gradient condition is thus crucial for the applicability of the perturbation approach. 
For GEP($k$) with maximal occupancy $k$, the gradient condition is obeyed in extreme cases of $k=1$ which reduces to the simple exclusion process and $k=\infty$ which reduces to random walks. Presumably because GEP($k$) is sandwiched between two extreme cases in which the perturbation approach works, this method provides a very good approximation when $1< k<\infty$.

We now clarify the underlying assumptions behind the perturbation approach and suggest some tracks to improve our results. For the GEP(2), the current reads
 \begin{align} 
 J_i = \langle \tau_i f (\tau_{i+1}) - f (\tau_i ) \tau_{i+1} \rangle ,
 \end{align}
where $ \tau_i \in \{0,1,2\} $ and $ f(n) =1 - \frac 1 2 n(n-1) $. 
In our computation of the diffusion coefficient \cite{we_14}, we used two assumptions. 
The first one concerns one-point functions. Let $\mathbb P [ \tau_i = m ]$ be the probability of finding $m$ particles at site $i$. The density at $ i $ is 
\begin{align}
 \rho_i = \langle \tau_i \rangle = \mathbb P [ \tau_i = 1 ] + 2 \mathbb P [ \tau_i = 2] .
\end{align} 
We assumed that one-site probabilities satisfy 
\begin{align} 
\label{eq:P=X}
 \mathbb P [ \tau_i = m ] \simeq X_m (\rho_i) \quad 
\end{align}
where the $ X_m $'s represent the single-site weights in an infinite lattice or on a ring: 
\begin{align}
 X_0 ( \rho ) = \frac{1}{Z}, \ X_1 ( \rho ) = \frac{\lambda }{Z}, \ X_2 ( \rho ) = \frac{\lambda^2}{2Z} 
\end{align}
with the fugacity $ \lambda $ and the normalization $ Z$ 
 \begin{align}
 \lambda ( \rho ) = \frac{ \sqrt{1+2\rho-\rho^2} + \rho -1 }{ 2-\rho }, 
 \ Z = 1 + \lambda + \frac 1 2 \lambda^2 . 
 \end{align}
The second assumption was to rewrite the current as 
\begin{align}
\label{eq:<>=<><>}
J_i \simeq \langle \tau_i \rangle \langle f (\tau_{i+1}) \rangle - \langle f (\tau_i ) \rangle \langle \tau_{i+1} \rangle .
\end{align}
This, indeed, is a mean-field type assumption \cite{Com}. The assumptions \eqref{eq:P=X}, \eqref{eq:<>=<><>} are asymptotically \textit{true} in the stationary state of a large system ($ L\to \infty $): We have checked these facts by performing additional simulations.

Our numerical results suggest more precise expressions for \eqref{eq:P=X} and \eqref{eq:<>=<><>} 
with some scaling functions $\kappa $ and $ \mu $: 
\begin{align} 
\label{eq:P=X+kappa}
 \mathbb P [ \tau_i = m ] = X_m (\rho_i) + \frac 1 L \kappa_m \Big( \frac i L \Big)\,, 
\end{align}
\begin{align}
\label{eq:<>=<><>+mu}
J_i =\langle \tau_i \rangle \langle f (\tau_{i+1}) \rangle -
 \langle f (\tau_i ) \rangle \langle \tau_{i+1}\rangle + \frac 1 L \mu \Big( \frac i L \Big) , 
\end{align}
where we omitted $o(L^{-1})$ terms. Performing the perturbation approach with the refined expressions \eqref{eq:P=X+kappa}, \eqref{eq:<>=<><>+mu}, we obtain 
\begin{align}
\label{J:exact}
 J = - \frac{1}{L} \frac{d\rho}{dx} 
 \left(1-X_2(\rho) + \rho \frac{ dX_2 ( \rho ) }{ d\rho } \right) + \frac 1 L \mu(x )
\end{align}
where we have switched from the discrete variable $ i $ to $ x= i/L $. The functions $ \kappa_m$ do not appear in \eqref{J:exact}, but $\mu(x)$ does, and it was missing in our paper \cite{we_14} leading to the wrong expressions for the current and for the stationary density profile. In order to calculate $\mu(x)$, we are presently examining nearest-neighbor correlation functions for the GEP(2). Numerically at least, these nearest-neighbor correlations exhibit a neat scaling behavior and simple patterns; detailed results will be reported in \cite{we_future}.


\begin{thebibliography}{99}

\bibitem{Com}
 T. Becker, K. Nelissen, B. Cleuren, B. Partoens, and C. Van den Broeck,
 Phys. Rev. E \textbf{93}, 046101 (2016). 
 
\bibitem{we_14}
 C. Arita, P. L. Krapivsky, and K. Mallick, Phys. Rev. E \textbf{90}, 052108 (2014). 

\bibitem{Spohn} 
 H. Spohn, {\it Large Scale Dynamics of Interacting Particles} (New York: Springer-Verlag, 1991).

\bibitem{KLS} 
 S.~Katz, J.~L.~Lebowitz, and H.~Spohn, J. Stat. Phys. \textbf{34}, 497 (1984).

\bibitem{Krapivsky} 
 P. L. Krapivsky, J. Stat. Mech. P06012 (2013). 

\bibitem{CCGS}
 J. M. Carlson, J. T. Chayes, E. R. Grannan, and G. H. Swindle, Phys. Rev. Lett. \textbf{65}, 2547 (1990). 

\bibitem{GK}
 D. Gabrielli and P. L. Krapivsky, in preparation.

\bibitem{C-T}
 C.~Cocozza-Thivent, Z. Wahrscheinlichkeitstheorie verw. Gebiete \textbf{70}, 509 (1985).

\bibitem{AM}
 C. Arita and C. Matsui, arXiv:1605.00917. 

\bibitem{KLO94} 
 C. Kipnis, C. Landim, and S. Olla, Commun. Pure Appl. Math. \textbf{47}, 1475 (1994).

\bibitem{KLO95} 
 C. Kipnis, C. Landim, and S. Olla, Ann. Inst. H. Poincar\'e \textbf{31}, 191 (1995).

\bibitem{Timo} 
 T. Sepp\"al\"ainen, Ann. Prob. \textbf{27}, 361 (1999).

\bibitem{BNCPV13}
 T. Becker, K. Nelissen, B. Cleuren, B. Partoens, and C. Van den Broeck, Phys. Rev. Lett. \textbf{111}, 110601 (2013).
 
\bibitem{we_future} C. Arita, P. L. Krapivsky, and K. Mallick, in preparation.
 

\end{thebibliography}
\end{document}